# High resolution 3D imaging of diamonds with multiphoton microscopy

ELANA G. ALEVY,[1,*] SAMUEL D. CROSSLEY,[1] LAM T. NGUYEN,[2] VU D. PHAI,[3] & KHANH Q. KIEU[2]

[1]*Lunar and Planetary Laboratory, University of Arizona, 1629 E. University Blvd., Tucson, AZ 85721, USA*
[2]*Wyant College of Optical Sciences, University of Arizona, 1630 E. University Blvd., Tucson, AZ 85721, USA*
[3]*Department of Electrical and Computer Engineering, University of Arizona, 1230 E. Speedway Blvd., Tucson, AZ 85719, USA*
*\*alevy@lpl.arizona.edu*

**Abstract:** Diamonds offer unique benefits for optical technology development due to their optical, chemical, electrical, mechanical, and thermal properties. These attributes also contribute to their aesthetic appeal, high commercial value, and utility in geological studies. Thus, there is high demand for nondestructive techniques that enable rapid analysis of natural and synthetic diamonds as well as diamond-like simulants. Here, we demonstrate sub-micrometer, nondestructive, three-dimensional imaging and spectral analysis of diamonds using multiphoton microscopy (MPM). This approach stimulates nonlinear optical emissions to provide unique insights into the interior structure, fluorescent defects, and formational conditions of diamonds. As a result, MPM can be used to investigate gemstone quality, vacancy centers used in quantum technologies, and the various inclusions and fluorescent emitters that may trace gemstone provenance and treatment history.

## 1. Introduction

Diamonds have a high refractive index (~2.4) and a broad transparency window (from UV to far-IR; 0.25 to 100 μm) [1]. They are also the hardest materials with high thermal conductivity and chemical inertness, which makes them suitable for a variety of technological applications [2]. When formed under specific environmental conditions, the optical properties of diamonds can be controlled for specific uses across commercial, industrial, and scientific applications (e.g. [3] and references therein).

Natural diamonds form deep in the earth (typically between 150 and 250 km [4]) at high pressures and temperatures. They can be classified based on the origin of their color, whether naturally occurring or artificially treated. In addition, natural diamonds can exhibit chemical impurities commonly due to the element nitrogen (N), which replaces carbon in the crystalline lattice. Type I diamonds exhibit abundant N defects that can be arranged in two ways: either replacement into nearby positions (Type Ia, including adjacent Type IaA or complex groupings in Type IaB) or in isolated lattice spots (Type Ib). The configurations of these nitrogen impurities vary depending on the host diamond's growth environment [5–9]. The most common defect exhibited by natural diamonds is the N3 nitrogen vacancy center produced from three nitrogen atoms surrounding a structural vacancy ($N_3V$). The N3 vacancy center exhibits characteristic emission around 415 nm, which has been identified as a diagnostic spectral characteristic of natural diamonds [10].

Synthetic diamonds and diamond-like simulant alternatives such as cubic zirconia ($ZrO_2$) and moissanite (SiC) exhibit unique chemical and structural characteristics that distinguish them from naturally occurring diamonds. Beginning with the first laboratory-synthesized diamond in 1955 [11], the chemical vapor deposition (CVD) or high-pressure-high-temperature (HPHT) processes have expanded the availability of gemstone-quality diamonds. Under vacuum with temperatures between 700ºC and 1300ºC, hydrocarbon gas and hydrogen are heated by an energetic source, which enables the release of carbon atoms onto a cool substrate plate, thus forming a CVD diamond. Characteristically, diamonds produced by CVD exhibit even clearness or coloration, contain few inclusions, and grow in tabular crystals. Growth occurs as discrete overlapping layers, the result of which are often distinguishable characteristics of CVD diamonds [12]. The growth of HPHT diamonds involves the heating of a carbon source and catalyst to pressures exceeding 5 GPa, the minimum bound for HPHT nucleation. In some applications, diamond seeds are included to control the geometry of HPHT-synthesized diamonds [3]. The diamonds produced from these synthesis methods vary in color, clarity, and size. As such, it is crucial to accurately distinguish between each of these diamond or diamond simulant materials using unique, quantitative benchmarks for each sample.

Current analytical techniques to classify diamonds commonly rely on the bulk-fluorescent signal emitted from a sample. These techniques have enabled reliable positive characterization of natural and lab-grown synthetic diamonds. For example, bulk fluorescence spectroscopy has enabled the measurement of NV3 and other nitrogen vacancy centers in diamonds (e.g. [10,13]). Raman spectroscopy has also been used to distinguish diamonds from diamond-like simulants. Finally, Fourier-Transform-Infrared (FTIR) spectroscopy is a powerful technique for diamond characterization that reveals characteristic nitrogen, hydrogen, silicon, and other defect centers [14].

However, these established techniques lack the capability to spatially resolve unique optical signatures indicative of defects, inclusions, or impurities in diamonds. It has been demonstrated that multiphoton microscopy (MPM) enables nondestructive and high-resolution imaging of geologic materials at all stages of sample preparation. MPM employs a pulsed femtosecond laser to deliver sufficient optical peak power into a sample such that multiple photons combine to excite electrons and endogenous fluorophores at double or triple the frequency of the incident laser. This process produces a variety of nonlinear optical signals including second harmonic generation (SHG), which is produced by phases lacking centro-symmetry in their crystalline structures. Third harmonic generation (THG) is produced at interfaces between materials with differing refractive indices. Two photon excitation fluorescence (2PEF), first demonstrated in biological samples [15], and three-photon excitation fluorescence (3PEF) can be produced from a variety of fluorophores in geologic materials. Common sources for 2PEF and 3PEF include trace abundances of rare earth elements (REE) and crystalline defects [16]. Recently, five-photon excitation fluorescence (5PEF) was imaged in a zoned fluorite sample [17]. However, MPM has not yet been systematically utilized to investigate diamonds. Here, we propose that multiphoton microscopy can be applied to produce high resolution, 3D images of the surface and interior of cut diamonds. In addition, we demonstrate that multiphoton microscopy can excite most defect centers in diamonds. This makes MPM a valuable tool for assessing gemstone quality, studying vacancy centers used in quantum technologies, visualizing various inclusions, and tracing fluorescent emitters that can record gemstone provenance and treatment history.

The first demonstrations of gemstone analysis with MPM focused on the utility of this technique in collecting subsurface textural data from a suite of terrestrial minerals. The nonlinear optical properties of minerals have been imaged over 600 microns beneath the surface of millimeter-scale cut and polished samples [18]. Additional MPM analysis of geologic materials has demonstrated that the distribution and type of nonlinear optical signal can uniquely highlight not only a geologic sample's subsurface structural properties but also sources of nonlinear fluorescence [19].

MPM analysis of diamonds has been limited to several applications including spectral analysis of two-photon fluorescence from multiphoton excitation of nitrogen vacancy centers and silicon vacancy ($Si-V^-$) centers in Si-implanted CVD diamond chips [20,21]. However, the classification and imaging of gemstone-quality diamonds have not yet been explored with multiphoton fluorescent and harmonic imaging. Here, we expand on previous diamond classification studies to demonstrate that MPM enables nondestructive, 3D mapping of the nonlinear optical properties of interior structural and chemical defects in cut natural diamonds, synthetic diamonds, and diamond-like simulant materials. This technique has broad implications for rapid and nondestructive gemstone imaging, quantitative analysis, and classification.

## 2. Methods

### 2.1 Multiphoton Microscopy

Multiphoton microscopy was performed on a custom-built microscope in the Ultrafast Fiber Lasers and Nonlinear Optics Group in the Wyant College for Optical Sciences at the University of Arizona (Fig. 1). A detailed design of the MPM system is documented by [22], and the setup is consistent with previous work on geologic materials [19]. The MPM contains a pulsed femtosecond 1040 nm laser with a 60 nm full-width-half-maximum (FWHM) bandwidth to stimulate nonlinear optical emissions. The laser produces 60 fs compressed pulses at a repetition rate of 8 MHz with ~50 mW of average power, which corresponds to ~6 nJ pulse energy. Due to energy loss along the optical path, only about ~30 mW is incident on the sample plane (~4 nJ).

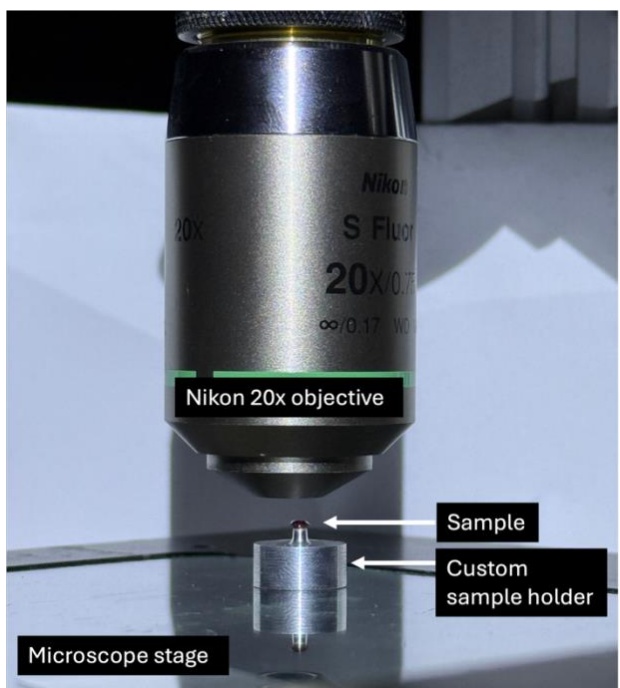

Fig. 1. Experimental setup for multiphoton microscopy of faceted gemstones, showing a synthetic CVD red diamond under the 20x microscope objective. A custom-built sample holder placed flat on the microscope stage enabled the table to be near-parallel to the objective. The diamond in this image is approximately 3 mm in diameter, but larger samples (cm-scale) can also be imaged by stitching multiple images together into a mosaic.

The following dichroic mirror and optical filter configurations enabled the simultaneous collection of two channels of signal in the photomultiplier tubes (PMTs): 1. SHG: 414 nm dichroic mirror and 517 nm bandpass filter (20 nm bandwidth); 2. THG: 414 nm dichroic mirror and 340 nm (26 nm bandwidth) bandpass filter; 3. 2PEF: 560 nm dichroic mirror and 594 nm longpass filter (or 750 nm shortpass filter); 4. 3PEF: 560 nm dichroic mirror and 447 nm bandpass filter (60 nm bandwidth). All channels had a pump filter (980 nm shortpass filter) to prevent the excitation light from reaching the photomultiplier tubes (PMTs).

Imaging was completed using either a Nikon 10X 0.5 NA or a Nikon 20X 0.75 NA objective. To compile three-dimensional scans for each diamond, two-dimensional images were collected as the sample was axially translated in steps until the full sample was imaged. At an imaging speed of 10 ms/line, the corresponding pixel dwell time was 10 μs/pixel. Scanning conditions for a depth profile typically included 10 μs/pixel dwell time, 500x500 or 1,000x1,000 pixels/frame, and either a 1 μm, 2 μm, 4 μm, or 5 μm vertical displacement step size between each image. The brightness and contrast for all MPM images were adjusted manually in ImageJ to set noise thresholds and enhance emission features for illustrative purposes.

Spectra of the samples were collected to obtain quantitative measurements of the imaged MPM signals. Using a cooled Ocean Optics QE6500 spectrometer, individual emission spectra were collected below the samples' surface to characterize the emission properties of inclusions, vacancy centers, and other sources of nonlinear optical signals in the SHG, THG, 2PEF, and 3PEF channels. Individual point spectra were obtained with a sampling volume on the order of single microns cubed. Representative spectra of the samples were collected beneath the top surface of the sample, where THG signal from the sample boundary interface was minimal. All point spectra were taken from either discrete inclusions or heterogeneities inside the sample. While it was also possible to collect spectra over an averaged field of view (100s of microns), point spectra were collected to limit uncertainty due to signal crosstalk between channels. Bulk spectra of large sampling areas provide representative spectral analysis, but the variation between nonlinear signals within each sample indicates that the point spectra method more accurately quantified the spectral characteristics of each sample.

Three dimensional datasets were processed using a custom artifact-removal and radius-interpolation algorithm in Python (Supplementary Algorithm 1). MPM imaging artifacts were produced due to total internal reflection at the polished facets of diamond samples. As a result, areas of MPM signal that mapped outside of the diamond sample were simply reflections of internal signal. THG signals were therefore used to identify the extent of real (i.e. non-artifact) MPM signal for each individual image in the 3D stack. The radii of circular regions of interest that fully enclosed the real MPM signals were identified for each image slice; these radii were then interpolated across slices to construct smooth, slice-consistent circular masks that remove unwanted background artifacts from all four channels. Other corrections included scaling the radius of the final slice and performing a linear interpolation of radii across the full depth profile to best reconstruct the diamond shape. Napari [23], an open-source, Python-based multi-dimensional image viewer enabled the display and manipulation of the three-dimensional datasets.

## 2.2 White light imaging

White light images of each sample were collected with the Keyence VHX-7000 Series digital microscope in the Lunar and Planetary Laboratory's Kuiper-Arizona Laboratory for Astromaterials Analysis (K-ALFAA) at the University of Arizona. Samples were placed with the culet embedded into a dark-colored sample holder with the table near-parallel to the microscope objective. Images were collected with varying focal distances such that both the sample's internal structure and the table geometry were visible.

## 2.3 Sample descriptions

Optical images and MPM analyses were collected for a range of cut diamonds and diamond simulants from various sources. Here, we present a suite of 8 representative samples including natural diamonds, synthetic diamonds, and diamond-like simulants (Fig. 2). The three natural diamonds include an Argyle diamond, a natural pink diamond, and a natural clear diamond. The Argyle diamond is the only sample studied here with a known provenance from the Argyle mine in Western Australia [24,25]. Other samples are from unknown provenances. Synthetic diamonds included both HPHT- and CVD-formed samples. The diamond-like simulants included moissanite and cubic zirconia. While samples were acquired from online gemstone vendors and their classifications should be considered provisional, we independently confirmed these classifications using Raman spectroscopy (Supplementary Fig. 1). The results of this work are therefore intended to demonstrate how multiphoton microscopy can be used to distinguish between various fluorescent defects and physical characteristics of diamonds using nonlinear optical emissions, thus providing a new tool to complement and enhance current analytical and appraisal techniques.

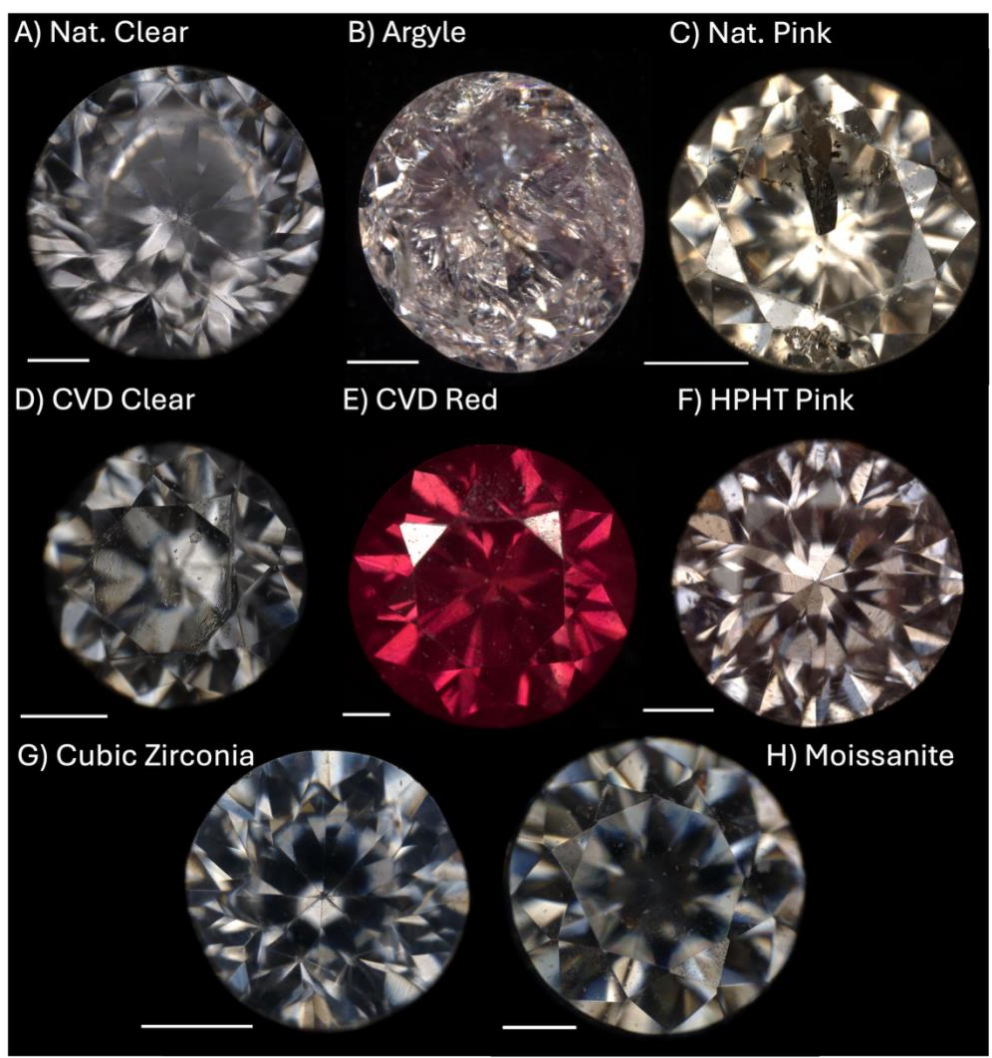

Fig. 2. White light images of natural diamonds (A-C), synthetic diamonds (D-F) and diamond-like simulants (G-H). Images were collected overhead and focused on the table facet. Scale bars are 500 µm.

## 3. Results

### 3.1 Natural diamonds

Coordinated analyses of white light imaging and multiphoton microscopy reveal the nonlinear optical response of heterogeneities in each natural diamond (Figs. 2-6, Table 1). In white light, the natural clear diamond lacks both color and resolvable inclusions near its subsurface (Fig. 2A). The Argyle diamond is light pink under white light conditions (Fig. 2B), and both optical and multiphoton imaging both reveal complex structural features throughout this sample (Fig. 5). The natural pink diamond (Fig. 2C) appears faintly yellow/pink in white light and contains a prominent ~500 μm-long inclusion near its surface.

Figure 3 shows a representative multiphoton dataset compiled through 3D analysis of the natural clear diamond in four colors, one for each multiphoton signal. Fluorescent inclusions are present throughout the interior of the clear diamond. They occasionally cluster in linear arrays (Fig. 4B) and are strongest near facets (Fig. 4D). Point spectra taken within fluorescent inclusions in the clear diamond show a complex of emission peaks centered around 450 nm, which is consistent with N3 centers ubiquitous in naturally formed type Ia diamonds [26]. Several μm-scale inclusions and defects are visible as dark dots within the optical image of the natural clear diamond, which is colorless in white light.

Multiphoton analysis of the Argyle diamond shows that planar features are present throughout the >1-mm depth profile as three-photon fluorescence and SHG signals (Fig. 5). A representative emission spectrum from the Argyle diamond (Fig. 5C) shows broad fluorescence with a peak centered around 400 nm. The natural pink diamond (Fig. 2C) exhibits similar emission spectra to the clear diamond, with a main peak centered again around 455 nm (Fig. 6). Depth profiling reveals that the large inclusion in the pink diamond exhibits both SHG and fluorescent emissions (Visualization 3). The pink diamond spectrum also exhibits localized THG in the interior, which is indicative a change in refractive index at the interface between diamond and a mineral inclusion.

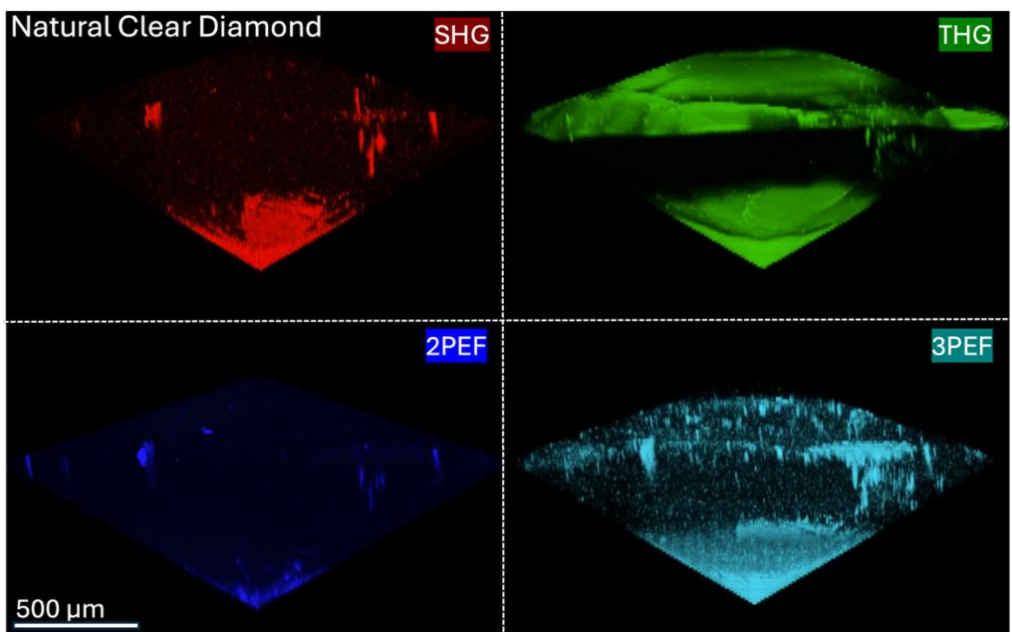

Fig. 3. Four false color multiphoton channels for the natural clear diamond, imaged parallel to the girdle, are rendered in 3D. SHG = second harmonic generation, THG = third harmonic generation, and 2,3PEF = two and three-photon excitation fluorescence, respectively. Harmonic signals are strongest at interfaces around the table, crown, and/or culet, as these facets are closer to perpendicular to the laser path. Fluorescent defects are present throughout the sample, and their signals are also amplified around the culet due to internal reflection. A composite, 4-channel rendering of the clear diamond is provided in the following figure. Animated videos of the two 3D renderings are provided in supplementary materials (Visualizations 8 and 9).

*3.2 Synthetic diamonds and diamond simulants*

White light imaging of the synthetic diamond samples shows the presence of small-scale inclusions and heterogeneities in all three synthetic samples (Fig. 5). The CVD red exhibits a deep red color, while the HPHT pink is a similar faint pink to the natural pink sample discussed previously. Multiphoton signals generated from synthetic diamond samples are shifted to longer wavelengths than natural diamond emissions (Fig. 6). The CVD red diamond exhibits strong fluorescence, mostly 2PEF, in the bulk of the diamond, which is quantified in the emission peak around ~660 nm. The CVD clear diamond exhibits strong emission at ~575 nm and ~650 nm with several resolvable peaks centered around 650 nm, characteristic of the $NV^{-1}$ center. The CVD red sample also contains several surficial heterogeneities and subsurface inclusions visible in MPM imaging (Visualization 5). The HPHT pink diamond emits most strongly around ~575 nm (characteristic of $NV^0$ centers) and exhibits the narrowest emission spectrum.

While cubic zirconia and moissanite appear similar to the clear natural and synthetic diamonds in white light, cubic zirconia lacks any substantial nonlinear optical signal except for THG (350 nm) at the sample edge. Moissanite only exhibits strong, ubiquitous second harmonic signal throughout its interior (525 nm). Additional imaging data for the synthetic diamonds and simulants are available in the supplementary material (Visualizations 1-9).

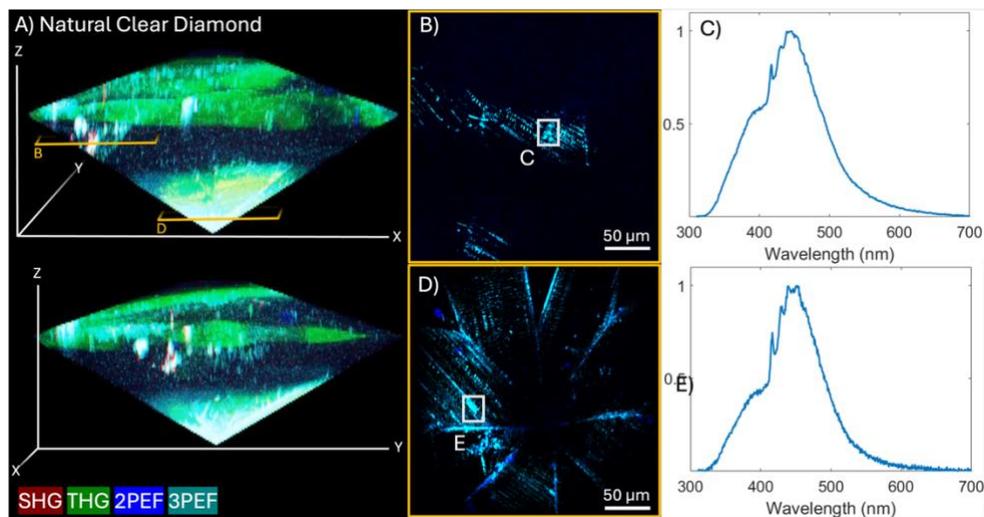

Fig. 4. False color MPM dataset for the natural Clear (CL-1) Diamond sample. The full volume renderings (A) show faint two-photon fluorescence (>530 nm) throughout the sample with hot spots of both two- and three-photon fluorescence dispersed throughout the interior. An interior slice ~150 µm below the crown highlights some of these hot spots, which are present as strings of µm-scale emitters in the highest resolution image (C). A bottom slice (D) shows radial bands of emission along facets near the point of the diamond. The spectrum of these emissions peak around 450 µm, which is consistent with N3 centers common in natural diamonds.

### 4.0 Discussion

*4.1 MPM as a diagnostic tool for high resolution 3D analysis of diamonds and other gemstones*

The nonlinear optical signals from diamonds and their gemological analogs may serve as a diagnostic tool for classification (Table 1). Specifically, the three-photon interactions stimulated at 347 nm (Fig. 6) are at comparable wavelengths to the 3$^{rd}$ order bulk fluorescent emissions stimulated by long-wave UV lamps conventionally used to classify diamonds [27].

The natural diamonds in this study exhibit characteristic blue fluorescence from ~400 nm to 450 nm. This wavelength range is consistent with N3 system defects, which account for ~97% of blue fluorescence observed in natural diamonds from previous works [28,29]. The pink natural diamond also contains a tail of fluorescence at longer wavelengths out to ~700 nm, which could be stimulated from NV centers. Overlapping fluorescence bands of unknown origin within this spectral range have often been measured in diamonds [27]. The high spatial resolution of MPM may allow the sources of these fluorescent bands to be directly investigated, but such analysis is beyond the scope of this work.

Synthetic diamonds fluoresce at longer wavelengths with broad peaks ranging from ~550–660 nm. Many of these signals are likely the result of overlapping fluorescence emissions from a variety of unidentified sources introduced during synthesis. However, MPM captures fluorescence from remaining NV centers–even in clear CVD diamonds. This is notable, as heat treatment of CVD diamonds (performed to increase clarify) often erase bulk fluorescence by reducing the number of NV centers during annealing [26]. Both CVD red and HPHT pink samples show sharp emission peaks at 575 nm, which is attributable to $NV^0$ defects [26]. Only CVD red shows a clear emission peak at 637 nm, indicative of the $NV^-$ defect used in quantum sensors [30].

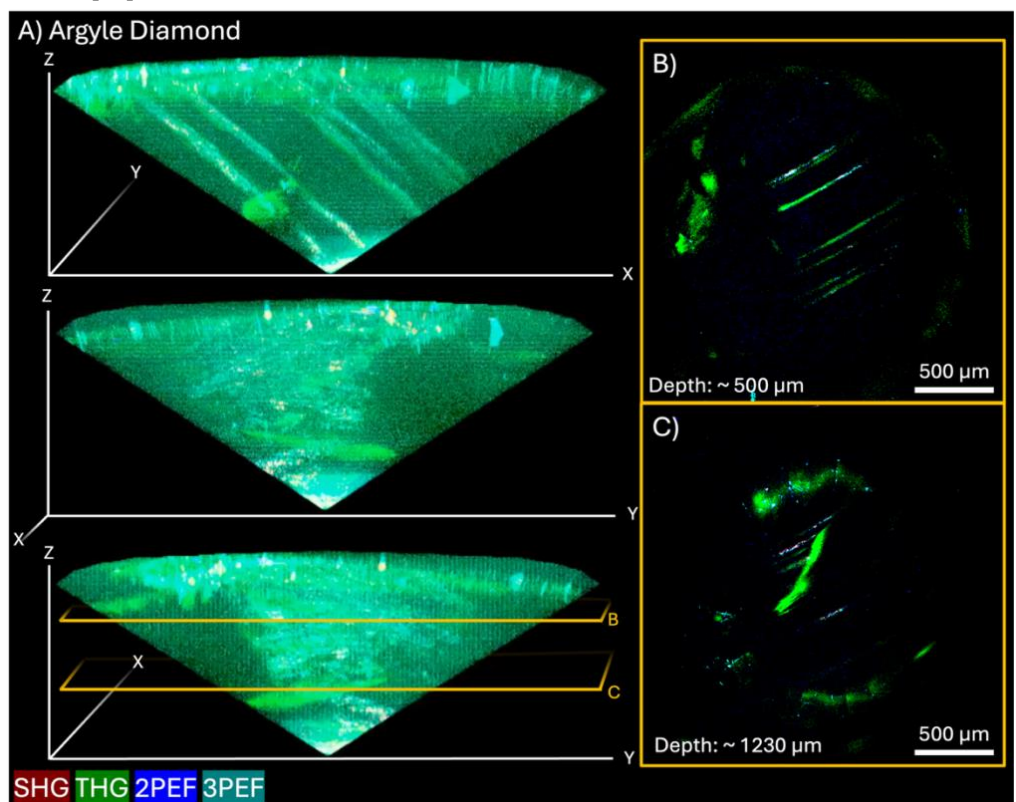

Fig. 5. False color multiphoton dataset for the Argyle diamond shows complex internal structures produced from plastic deformation during its residence in the Earth's crust. Parallel lamellae with overlapping THG, SHG, and 3PEF (B-C) show that shortwave fluorescent defects, likely attributable to N3 centers (Fig. 6), are concentrated in planar distortions. Such information not resolvable in other analytical methods and may be used to identify the provenance of certain samples. The 3D rendering is projected perpendicular to the laser path, parallel to the table facet, and rotated about the z-axis to highlight planar features.

The high spatial resolution of MPM provides unequalled 3D petrographic context for textural analysis of the diamonds. For example, the Argyle diamond exhibits planar structures in SHG that can be inferred as structural defects. SHG occurs in phases where crystalline symmetry is broken, so these planar features in the Argyle diamond are likely the nonlinear signature of strain due to plastic deformation in the diamond lattice. This is consistent with the overlapping THG signals along planar structures, which are produced where the refractive index changes across the focal volume of the laser. In gemological studies of diamonds, such structural features are conventionally investigated using birefringence in cross-polarized white light microscopy [24], a technique that also highlights changes in refractive indices. The diminished fluorescence signatures in Argyle may similarly be due to distortion of NV centers during plastic deformation.

Both moissanite and CZ lack the fluorescence observed in all natural and synthetic diamonds analyzed in this work and in other fluorescence studies [31]. These diamond-like simulant materials evidently do not form fluorescent emission centers during synthesis, though inclusions within them could still produce unique signals. Both moissanite and CZ exhibit THG at their surfaces due to high refractive indices. Moissanite also exhibits a strong SHG signal throughout the bulk, which is consistent with observations of other silicon carbide in geologic and manufactured materials [32,33]. The strong SHG in moissanite, as well as its high thermal conductivity and hardness, contribute to its effectiveness as a frequency doubling medium in optical technologies [33]. As bulk fluorescence is a measure commonly used to distinguish between diamond types, the ability for MPM to accurately distinguish between these diamond-like simulants in the absence of fluorescence is powerful.

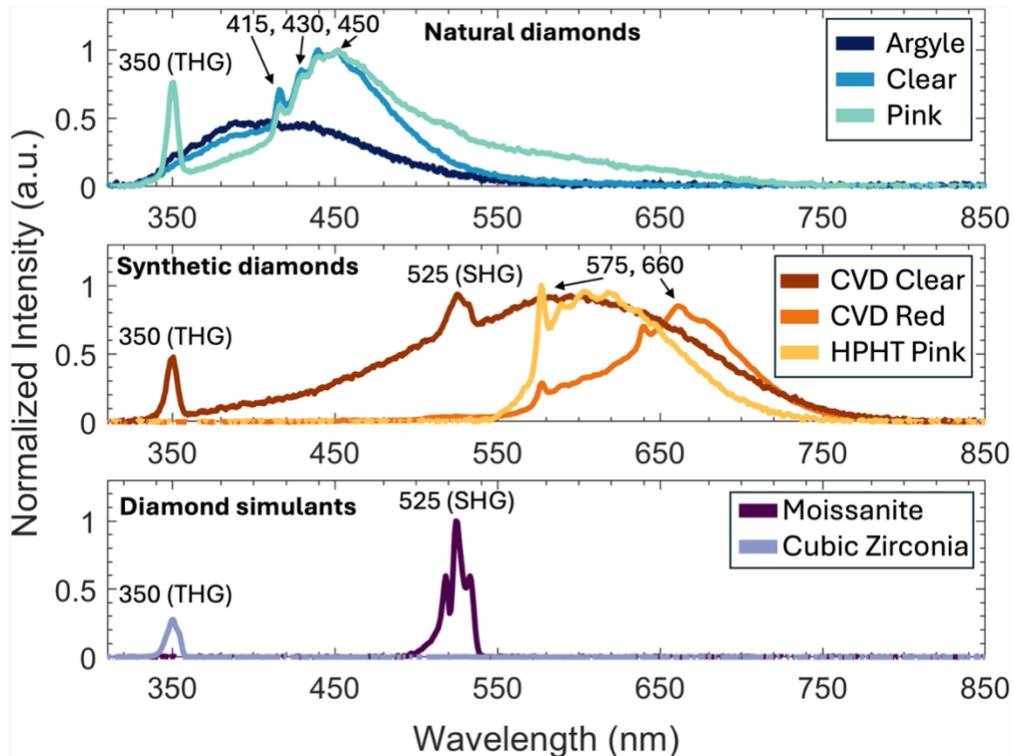

Fig. 6. Representative nonlinear optical emission spectra from diamonds and diamond-like gemstones show clear differences. Natural diamonds are typically characterized by 3PEF (~350-500 nm) with lesser 2PEF at longer wavelengths. These broad emissions are generally associated with N3 center complexes in the diamond structure and are characteristic of natural diamonds, as accumulation of N3 centers is a slow process. Synthetic diamonds fluoresce at longer wavelengths with broad central emission peaks from ~550-650 nm. Sharp emission peaks are evident in HPHT Pink and CVD Red, consistent with Nitrogen vacancy centers of various charges. CVD Clear lacks sharp peaks, possibly due to thermal annealing after synthesis to improve clarity. Diamond-like gemstones show no fluorescence in their interiors, but moissanite produces strong, ubiquitous SHG signals that can be used for identification.

*4.2 Comparison with other methods of diamond fluorescence analysis*

The only reported method for 3D fluorescent analysis of gemstones is hyperspectral photoluminescence [34]. While this approach allows for lateral spatial resolution of 1-3 μm, its axial resolution of ~35 μm/pixel (under the most ideal conditions) rapidly degrades with depth in the sample. That technique employs CW lasers that stimulate linear optical emissions along the laser's path, effectively limiting its utility in spatially resolving fluorescent emissions in three dimensions. In comparison, the MPM capabilities presented here provide axial resolution of 3-5 µm in diamonds and lateral resolution < 1 µm for both fluorescent and harmonic signals.

Multi-excitation photoluminescence, which can alternate between shortwave UV (261 nm) and UV-visible (405 nm) CW laser excitation, has recently proven useful for distinguishing between diamond types [35]. This technique has been used to measure fluorescent emissions from 270–1000 nm in diamonds, and it provides a means to reliably characterize diamond types from a broad range of optical emissions. Notably, the 2$^{nd}$ and 3$^{rd}$ order nonlinear optical emissions stimulated by MPM can access a similar range of fluorescent centers. The current MPM assembly described here detects emissions from ~300 nm to 1000 nm, but deep UV can be reached using a modified assembly including a shorter wavelength femtosecond laser and appropriate detectors.

The most widely used imaging technique for diamonds utilizes deep UV excitation to provide high resolution, 2D fluoresce images; emission spectra are usually collected from bulk samples cooled with liquid nitrogen [36]. Meanwhile, MPM produces fluorescence spectra and images in 3D with sub-micrometer spatial resolution. As a result, MPM advances the current state of analysis by providing a means to investigate fluorescence centers and crystallographic structures in situ, which is an important capability for prepared gemstones and manufactured materials–especially those with high value. In practice, this allows for the 3D petrographic analysis of diamonds and other gemstones using nonlinear optical emissions, as demonstrated with the Argyle diamond. Despite its nondestructive nature, photoionization of nitrogen vacancy centers may occur during multiphoton excitation. For example, $NV^0$ photoionizes to $NV^-$ at a wavelength of 532 nm [26,37]. Further study is necessary to fully characterize the effects of the MPM laser on the measured NV centers before quantifying the abundance of various defects.

Table 1. Nonlinear Optical Emissions from Diamonds and Simulants

| Samples | | Harmonics | | Fluorescent Defects | | |
|---|---|---|---|---|---|---|
| | | SHG | THG | N3 | NV0 | NV- |
| Natural | Clear | None | Facets | Well defined | None | None |
| | Argyle | Planar defects, facets | | Diffuse | None | None |
| | Pink | Mineral inclusions, facets | | Well defined | Weak | Weak |
| Synthetic | CVD Clear | Inclusions, facets | | Weak | Weak | Well defined |
| | CVD Red | Facets | Facets | None | Broad, poorly defined | |
| | HPHT Pink | Facets | Facets | None | Well defined | Weak |
| Simulant | Moissanite | Strong, uniform | Facets | None | None | None |
| | Cubic zirconia | None | Facets | None | None | None |

In summary, MPM achieves novel analyses that complement and improve upon the current analytical methods for gemstone analysis. Fluorescent emissions from 2$^{nd}$ and 3$^{rd}$ order nonlinear excitation expand on these analytical techniques by providing the following: (1) 3D image resolution due to the small focal volume in which nonlinear interactions occur, (2) improved signal-to-noise at comparable laser power output due to the exponential relationship between laser power and emission intensity [38], (3) structural analysis using harmonic generations stimulated in regions with broken symmetry and/or asymmetric electronic configurations, and (4) spatially-resolved fluorescence spectra of emissions from the interior of diamond samples. These unique capabilities allow for next-generation investigations into the sources of fluorescent emissions and provide a new suite of tools for the appraisal of gemstone quality and provenance.

## 5. Conclusions and future outlook

Multiphoton microscopy provides a high resolution, nondestructive approach for the 3D imaging of diamonds using nonlinear optical emissions stimulated by a femtosecond laser.

Harmonic generations and nonlinear fluorescent emissions provide new insight into the optical properties of structural and compositional variations in diamonds–these features can aid in diamond classification. Moreover, the high spatial resolution MPM (on the order of hundreds of nanometers) resolves crystallographic occurrences of emission centers that have diverse technological and scientific applications.

With continued development, this approach can contribute to novel appraisal techniques and determination of sample provenance for a wide variety of gemstones. Given its modular design, the MPM assembly can be configured to incorporate shorter wavelength femtosecond lasers capable of reaching the deep UV (<225 nm). Deep UV excitation stimulates fluorescent emissions throughout the bulk of most diamonds, which highlights growth patterns and other structural features indicative of synthesis methods. Alternatively, higher order multiphoton interactions could stimulate deep UV fluorescence [17], provided the assembly includes detectors sensitive enough to measure 5-photon emissions. Additional modification to the MPM could include a Raman spectrometry system that would further enable in situ verification of gemstone identity. Dichroic mirrors can also be selected to isolate signals from specific fluorescent centers, allowing for live image acquisition of defects including NV0, NV$^-$, and N3. These modifications illustrate the adaptability and suitability of MPM for targeted investigation into the formation history of diamonds and other gemstones.

**Disclosures.** The authors declare no conflicts of interest.

**Data availability.** Data underlying the results presented are not publicly available at this time but may be obtained from the authors upon reasonable request.